\documentclass[english,aps,reprint,10pt,onecolumn]{revtex4}
\usepackage[caption=false]{subfig}
\usepackage[T1]{fontenc}
\usepackage{color}
\usepackage{textcomp}
\usepackage{graphicx}
\usepackage{subscript}
\usepackage{babel}
\usepackage{amsmath,amssymb,amsfonts,hyperref}
\usepackage{epsfig,slashed,placeins}
\usepackage{braket,latexsym,epstopdf,rotating}
\usepackage[export]{adjustbox}
\usepackage{xcolor}
\usepackage[utf8]{inputenc}

\begin{document}
\title{
Solution of the two-center Dirac equation with 20-digits precision using the finite-element technique
}
\author{O. Kullie}
\affiliation{Theoretical Physics, Institute for Physics,
 Department of Mathematics and Natural Science, Universit{\"a}t Kassel, 34132 Kassel, Germany}
\email{kullie@uni-kassel.de}
 
\author{S. Schiller}
\affiliation{Institut f\"ur Experimentalphysik, Heinrich-Heine-Universit\"at D\"usseldorf,
40225 D\"usseldorf, Germany}
\email{step.schiller@hhu.de}
\begin{abstract}
We present a precise fully relativistic numerical solution of the two-center Coulomb problem. 
The special case of unit nuclear charges is relevant for the accurate description of 
the ${\rm H}_2^+$ molecular ion and its isotopologues, systems that are an active experimental topic. 
The computation utilizes the 2-spinor minmax approach and the finite-element method. 
The computed total energies have estimated fractional uncertainties of a few times $10^{-20}$ 
for unit charges and a bond length of 2 atomic units. 
The fractional uncertainty of the purely relativistic contribution is $1\times10^{-17}$. 
The result is relevant for future precision experiments, whereas at present the uncertainties 
arising from the quantum electrodynamic treatment of the rovibrational transition frequencies 
are dominant.  
\end{abstract}
\maketitle
\nopagebreak
\section{Introduction}\label{Int}
There is currently considerable interest in precisely measuring the rotational and vibrational 
transitions in the molecular hydrogen ions and comparing the values with {\it ab initio} theory predictions \cite{Korobov2021}. 
Such comparisons allow a series of applications: determination of mass ratios, of nuclear charge radii, 
tests of wave mechanics, and search for fifth forces \cite{Alighanbari2020,Kortunov2021,Patra2020}. 
Disregarding hyperfine structure contributions, the relativistic contributions are the largest ones beyond 
the Schr{\"o}dinger energy. 
They make an important, easily measurable, contribution to rotational and vibrational transition frequencies 
of the molecular hydrogen ions. For example, for the overtone vibrational transition 
$(v=0,L=0)\rightarrow(v'=5,L'=1)$ of ${\rm HD}^+$ this contribution is $\simeq4.0\,$ GHz, or $1.5\times10^{-5}$ 
relative to the transition frequency. Here $v,\,L$ are the vibrational and rotational quantum numbers of the level. 
Today's experimental uncertainty of this transition frequency is of the order of 1\,kHz, i.e. $2.5\times10^{-7}$ 
of the relativistic contribution , and approximately $1\times10^{-11}$ relative to the transition frequency 
itself \cite{Alighanbari2021}. There are excellent prospects for further reduction of the experimental 
uncertainty in the near future: this would be achieved using techniques already demonstrated for 
the precision spectroscopy of single atomic ions, that have reached uncertainties below the $1\times10^{-17}$ 
level (see Ref.\,\cite{Ludlow2015} for a review). Two aspects can be mentioned in this respect. 
First, the controlled trapping of single molecular hydrogen ions has recently been 
demonstrated \cite{Wellers2021} and, second, the systematic shifts of vibrational transitions have been 
analyzed theoretically and found to allow an uncertainty at the level below $1\times10^{-16}$ \cite{Schiller2014}. 
Thus, it is clearly desirable to perform a highly precise theoretical evaluation of the relativistic contribution. 

The currently employed approach to deal with the dominant relativistic effects is the perturbative 
evaluation of the Breit-Pauli Hamiltonian, with respect to the electronic wave function computed in 
the approximation of fixed nuclear charge centers \cite{Tsogbayar:2006}. 
This gives the relativistic shift of the order of $\alpha^2$ compared to the nonrelativistic energy, the latter 
being close to $-1$ atomic unit (a.u.). Beyond this, the shift of relative order $\alpha^4$ can also be 
computed perturbatively, with an appropriate formalism \cite{Korobov:2007}. Finally, computations of the 
contribution of the order of $\alpha^6$ have been available since the late 1980s.

In order to verify and extend the perturbation results, here we perform a high-precision numerical solution 
of the Dirac equation for the electron in the field of two static positive charges. 
We reduce the uncertainty of the relativistic shifts by a factor $>1\times10^7$ compared to the best previously 
published finite-element method (FEM) calculation, from $1\times10^{-14}$\,a.u. \cite{Kullie:2001} to below $1\times10^{-21}$\,a.u. 

The paper begins in Sec. II A with a brief introduction of the minimax approach for finding solutions of 
the Dirac equation, circumventing the issues found in other approaches. 
We also explain the iteration procedure and the nonrelativistic limit (Sec. II B). 
The implementation is discussed in Secs. II C and D.
Section III presents the computational aspects, including convergence of the FEM calculations and 
the treatment of the limiting cases of the hydrogen atom and of the nonrelativistic molecular hydrogen ion.  
The main results are contained in Sec.\,IV, viz. the relativistic shift as a function of 
the fine-structure constant and as a function of distance between the charge centers. Finally, 
Sec.\,V evaluates the consequences of the present treatment of the relativistic shifts on transition 
frequencies already measured for the molecular hydrogen ion.

\section{Method}\label{Method}
We apply the method  previously developed in the works of one of us (O.K.) \cite{Kullie:2001,Kullie:2003,Kullie:20041}. 
A minmax principle is used that is based on the elimination of the small component from the Dirac equation, 
leading to a non-linear eigenvalue problem that is solved iteratively.
The main extension implemented is this work is the use of larger FEM order, larger number of grid points, 
and the use of quadruple precision (32 digits) in order to achieve a high numerical accuracy. 
Also, stringent tests are performed that allow one to determine the uncertainty of the energy values obtained 
in the numerical solution. Specifically, we obtain a highly accurate result for the {${\rm H}_2^+$} system.
We also show that the chosen relativistic treatment has an efficiency approaching that of the solution of 
the non-relativistic Schr\"odinger equation.

\subsection{Concept}\label{concept}

A solution of the one particle 4-spinor Dirac equation can be obtained from a stationarity principle 
for the functional  $I=\langle\psi|\hat{H}|\psi\rangle-\varepsilon\langle\psi|\psi\rangle$ but one cannot 
apply a variational minimum principle as for the Schr{\"o}dinger equation. 
However, there exists a minimum principle in the space of bound electronic states.
The minmax principle, see \cite{Dolbeault:20002} (and references therein) \cite{Talmann:1986}, 
applies to the construction of the eigenvalues of an operator $ \hat H$ that has a gap in its 
continuous spectrum (here from $-m c ^{2}$ to $m c^{2}$) and that is unbounded from above and below. 
The principle considers the subspace of positronic states, $F_-$ and the subspace of electronic 
states, $F_+$ and requires a two-step search for extrema.
The sequence of minmax level energies is given by 
\begin{equation}
\label{eq1} 
\lambda_{k}=\inf_{\stackrel{{\rm dim} G=k}{G \mbox{\footnotesize{ subspace of }} 
F_{+} \,\,\,}} \sup_{\stackrel{\psi\neq |0\rangle}{\,\,\, \psi \in (G\oplus F_{-})}}
\frac{ \langle \psi \mid \hat H \mid\psi \rangle}{ \langle  \psi\mid  \psi\rangle}\ ,
\end{equation}
\noindent
where $F_{+}\oplus F_{-}$ is an orthogonal decomposition of a well-chosen 
space of smooth square integrable functions and 
${\langle \psi \mid \hat H \mid\psi \rangle}/{\langle  \psi\mid  \psi \rangle}$ 
is the Rayleigh quotient.
It has been proven \cite{Dolbeault:20002} that the sequence of the minmax energies $\lambda_{k}$ 
equals the sequence of eigenvalues of $\hat H$ in the interval $(-m c^2, +m c^2)$.  
The minmax principle transforms the problem of finding a solution of the Dirac equation to 
a minimization (infimum) problem. It guarantees a solution of the Dirac equation in the space of 
the large component  $\phi^{+}$. 
The spectrum consists  only of positive  eigenvalues, i.e. the  negative eigenvalues are 
eliminated and the spectrum is free from spurios states \cite{Dolbeault:20002,Talmann:1986}.
\subsection{The minmax eigenvalue equation}\label{minmaxf}
The Dirac eigenvalue equation of the electron in a scalar potential $V$,
$\widehat H_{D}\, \psi= \varepsilon \psi$, with the 4-spinor $\psi$, can be written in the form 
\begin{eqnarray}\label{eq:Diraceq}
 \left(  
\begin{array}{cc} V & \hat L^{\dagger}  \\ 
                 \hat L\ &  V- 2 m c^2
\end {array} \right)
\left( \begin{array}{c}
  \phi_{+}\\
  \phi_{-}
\end{array} \right)   
=  \varepsilon  \left( \begin{array}{c}
  \phi_{+}\\
  \phi_{-}
\end{array} \right)\ , 
\end{eqnarray}
where we introduced the 2-spinors $\phi_{+}$ and  $\phi_{-}$ for the large and small 
component of $\psi$, respectively. 
{$\hat L=-i \, c \, \hbar \, {\bf{\sigma}} \cdot \nabla$,
${\bf \sigma}=\sum_{k=1}^{3}\sigma_k\, e_{k} $}, where $\sigma_k$ are the Pauli matrices.
$\varepsilon$ is the eigenenergy that in the non-relativistic limit corresponds to the eigenenergy of 
the Schr\"odinger equation. 
Rather than solving Eq. (\ref{eq:Diraceq}) directly, or operating with the functional $I$ above, 
we proceed to reduce the 4-spinor treatment to a 2-spinor treatment. First, we eliminate the small 
component $\phi_{-}$ from Eq. (\ref{eq:Diraceq}), obtaining the differential ("strong") form 
\begin{equation} 
\label{eq:LdaggerL}
\hat L^{\dagger}\left(\frac{\hat L \, \phi_{+}}{\varepsilon + 2 m c^2  -V}\right) 
= (\varepsilon - V)\,\phi_{+}\ .
\end{equation}  
In addition, we turn to a "weak" (integral) formulation that provides a good efficiency for FEM with  
large finite-element basis sets. It is obtained by multiplying both sides in Eq. (\ref{eq:LdaggerL})
with $ \phi_{+}^{\dagger}$ and integrating over the electron's coordinate space \cite{Dolbeault:20002}: 
\begin{equation}\label{eq:integralform}
\int  \frac{| \hat L\phi_{+}|^2}{\varepsilon + 2 m c^2 -V} dr^3 = \int (\varepsilon  - V)\,|\phi_{+}|^2dr^3\ .
\end{equation}
We now apply the minmax principle and seek the minimum value of $\varepsilon$. 
We expand the 2-spinor $\phi_+$ over a set of basis functions with unknown coefficients. 
This set should be as large (complete) as possible. The vector of expansion coefficients 
will be denoted by ${\overline{\phi}}_+$.  
Variation of  Eq. (\ref{eq:integralform}) with respect  to the unknown coefficients in combination 
with the requirement that $\varepsilon$ is stationary with respect to all coefficients 
leads to a matrix equation that determines the coefficients. It reads 
\begin{equation}\label{eq:matrixeq}
{\bf M }(\varepsilon)\, {\overline{\phi}}_+ = \varepsilon \,{\overline{\phi}}_+\ .
\end{equation}
 Thus, the original 4-spinor Dirac equation has been transformed into a Schr\"odinger-like equation 
 for the 2-spinor $\phi_+$, but where the effective "Hamiltonian" ${\bf M}$ is eigenvalue-dependent. 
 The equation is nonlinear in the eigenvalue $\varepsilon$ and therefore has to be solved by iteration.
It can be shown that the solutions  of this equation minimize the Rayleigh quotient over all 
electronic bound states of $\hat{H}_D$ \cite{Dolbeault:20003}.

It has been shown that an efficient approach consists of expanding the left-hand side of 
Eq. (\ref{eq:integralform}) in a series \cite{Dolbeault:20002}, as follows.
We start with an approximate value $\varepsilon^0$ of an eigenvalue $\varepsilon$. 
For the iteration $j+1$ $(j=0,...,j_{\rm max})$ we expand the left-hand side as  \cite{Kullie:20042}:
\begin{eqnarray}\label{eq:E0jjexpansion}
&&\int\frac{| \hat L\phi_{+}|^2}{\varepsilon^{j}+ 2 m c^2 -V} dr^3=
\int\frac{| \hat L\phi{+}|^2}{\underbrace{(\varepsilon_0 + 2 m c^2 -V)}_{g(\varepsilon_0)}+ 
\Delta \varepsilon^{j}}dr^3= \\\nonumber
&&\int\frac{|\hat L\phi_{+}|^2}{g(\varepsilon_0)}dr^3
+\sum_{k=1}^{k_{\rm max}} (-\Delta \varepsilon^{j})^{k} \int\frac{|\hat L\phi_{+}|^2}{g(\varepsilon_0)^{k+1}} dr^3\ ,
\end{eqnarray}
with $\Delta  \varepsilon^{j} = \varepsilon^{j} - \varepsilon_0$. 
The series expansion has the advantage that the iteration procedure reduces to solving successive eigenvalue problems. 
At iteration ${j+1}$ one computes the updated global matrix corresponding to the expression 
(\ref{eq:E0jjexpansion}), ${\bf M'}(\varepsilon_0,\Delta\varepsilon^{j})$.  
One then solves the conventional eigenvalue problem 
${\bf M'}(\varepsilon_0,\Delta\varepsilon^{j}){\overline\phi}_+=\epsilon^{j+1}{\overline\phi}_+$. 
This is the computationally heaviest and therefore longest part of the numerical procedure. 
Another advantage of the series expansion is that the matrix elements for the individual terms 
in  (\ref{eq:E0jjexpansion}) need to be computed just once for a given grid, at the beginning of the iteration. 
For explicit expressions of the elements of the matrix Hamiltonian ${\bf M'}$ we refer to Ref.\,\cite{Kullie:20042}.

The matrix  equation is solved by an iterative method with a Cholesky decomposition \cite{Heinemann:1987}. 
In our method only an open boundary condition is implemented so far, which has been found to work well.

The iteration process is stopped at $j={j}_{\rm max}$ or when a required accuracy between successive 
iterations is reached. The series converges quickly and for atoms with small nuclear charge only 
$k_{\rm max}=3$ to $4$ terms are needed. Typically, a small number of iterations is sufficient 
$j_{\rm max}=3$ for the $Z=1$ case. 
For large-$Z$ nuclei, usually $k_{\rm max}=4$ to $9$ terms are sufficient 
and $j_{\rm max}=3-7$ remains small or moderate.
For the nonrelativistic case  $c\longrightarrow \infty$ the $k$ expansion is unnecessary, see below. 

In the FEM approach, one usually performs the computation for a series of grids with increasing 
number of elements and thus increasing number of basis functions. When one moves from one grid to 
the next finer grid, one uses as new start value $\varepsilon^0$ the solution 
$\varepsilon^{j_{\rm max}+1}$ found in the previous grid.

The 2-spinor formulation exhibits major advantages compared to the 4-spinor formulation: 
the number of  matrix elements to be computed is a factor 3 smaller and the solution of 
the matrix (by inverse vector iteration)  
requires a factor of 4 fewer operations. The reduced size of the problem enhances the computational
performance and allows one to tackle larger problems \cite{Kullie:20041,Kullie:20042}.

Note that Eq. (\ref{eq:integralform}) can be written in the form  
\begin{equation} \label{eq:nrlimit} 
\int{\frac{c^{-2}| {\hat L} \phi^{+}|^2}{(2 m+(\varepsilon -V)/c^2)}} 
= \int (\varepsilon  - V)\,|\phi_{+}|^2dr^3\ .
\end{equation}
Therefore, in the nonrelativistic limit ($c\to \infty$), Eq. (\ref{eq:integralform}) turns into 
the Schr\"odinger equation, considering that $\hat{L}$ is proportional to $c$. 
Thus, we recognize that Eq. (\ref{eq:integralform})
exhibits similar properties to the Schr\"odinger equation.
In practice, we calculate the nonrelativistic values by setting $c$ to a large number, 
$c\ge 10^{15} (\alpha^{2} \le 10^{-30})$. 
In this limit the small component $\phi_-$ becomes zero and the two components of 
$\phi_+$ are then identical. 
The possibility of computing the nonrelativistic energy value in this way 
(i.e. using the same numerical procedures) leads to an important advantage: 
by subtracting it from the value for finite $c$ we can extract the  relativistic 
{\em shift} with a better accuracy than the accuracy of the total 
(nonrelativistic plus relativistic shift) energy (see Sec.\, \ref{RaD}). 

We showed in previous work \cite{Kullie:20041,Kullie:20043} that in the weak formulation  the 2-spinor fully 
relativistic FEM approach to the two-center Coulomb problem is numerically 
better behaved than the numerical solution of the 4-spinor Dirac equation. 
One finds some very desirable behaviors: 
the energy values converge from above with increasing grid size (finer 
subdivisions) and do not show the typical convergence from below  or 
oscillatory convergence of the 4-spinor Dirac equation.  
This is a consequence of the elimination of the small component $\phi_-$, 
which effectively projects the problem onto electronic states and leads to a second-order 
differential operator bounded from below.  
\subsection{Implementation}\label{Met}
The Dirac Hamiltonian for a single particle of mass $m$ in a two-center potential $V$ is 
\begin{eqnarray}\label{eq:hhD} 
\widehat H_D = c \,\widehat{\alpha}
               \cdot \widehat{\bf{p}} + m_{\rm} c^2 \widehat{\beta} +V, \\ \nonumber
\mbox{} V = - \sum_{l=1}^2 \frac{\hbar\, c\, \alpha\, Z_l}{|{\bf r} - {\bf R}_l|}\ .
\end{eqnarray}
$Z_{1}, Z_{2}$ are the charges of the two nuclei in units of the elementary charge, 
$\widehat{\bf \alpha}$ and $\widehat\beta $ the usual Dirac matrices, ${\bf r}$ is the position 
of the electron, $\widehat{\bf{p}}$ is the momentum operator, and ${\bf R}_l$ are the positions of the nuclei. 
$\alpha$ is the fine-structure constant.
Alternatively, if atomic units are employed,
\begin{eqnarray}\label{eq:hhD2} 
\widehat H_D &=& m_{\rm} c^2 \alpha^2
\left(\alpha^{-1}
\widehat{\alpha}\cdot \widehat{\bf{p'}} + 
\alpha^{-2}\widehat{\beta} + V'\right)\ , \\ \nonumber
\mbox{} V' &=& - \sum_{l=1}^2 \frac{Z_l}{|{\bf r}' - {\bf R'}_l|}\ .
\end{eqnarray}
The primed quantities correspond to the case when coordinates and momenta are in atomic units.   
$m_{\rm}\alpha^2 c^2$ is the atomic unit of energy. 
  
The nonrelativistic energy is found from the difference between the total energy and 
the rest-mass energy $m_{\rm}\,c^2$, in the limit $c\rightarrow\infty$ of Eq. (\ref{eq:hhD}), 
or equivalently, in the limit $\alpha=e^2/(4\pi\epsilon_0\hbar\,c)\rightarrow0$ of Eq. (\ref{eq:hhD2}). 
In both cases, the product $c\,\alpha$, i.e. the potential energy $V,V'$  and the atomic energy unit, 
are to be kept constant.  In the following, often $c$ and $\alpha^{-1}$ are used synonomously; 
they are equal in an appropriate system of units. For the atomic case ($Z_2=0$), we consider different 
values of $Z_1$, in particular $Z_1\gg1$; see below.

For the two-center case one has axial symmetry around the internuclear axis (the $z$-axis) and  
favorably uses prolate spheroidal (elliptic spheroidal) coordinates $\xi $ and $\eta $,
\begin{eqnarray}
x=\frac{R}{2} u(\xi,\eta) \, \cos\varphi, \,y=\frac{R}{2} u(\xi,\eta) \, \sin\varphi, \\\nonumber
z=\frac{R}{2}\xi\cdot \eta, \mbox{ where } u(\xi,\eta)=\sqrt{(\xi^{2}-1)(1-\eta^{2})}\,\
\end{eqnarray}
and $R$ is the inter-nuclear distance in atomic units.  
The electron's angular coordinate is $\varphi$. 
The distances between the electron and the nuclei are
\begin{equation}\label{eq:xieta}
    r_{1}=(\xi+\eta) \frac{R}{2},\quad r_{2}=(\xi-\eta) \frac{R}{2}\ .
\end{equation}
The Coulomb singularity of point nuclei causes a singular behavior 
of the relativistic solutions at the position of the nuclei of the form 
$r^{-1+\gamma_{l,\kappa}}_{l}$, with  $\gamma_{l,\kappa}=\sqrt{\kappa^{2}-
{ (\alpha Z_{l})^2}}$ and $\kappa=|j_{z}|+\frac{1}{2},\, l=1,2$. 
This is well-known from atomic calculations \cite{Yang:1993,Duesterhoeft:1994}. 
Thus, further singular coordinate transformations (whose back transform is non-analytic) 
are needed to take care of this issue \cite{Kullie:2001,Kullie:20042,Kullie:2003}. 
The transformation from $\xi,\eta$ to $s,t$ reads
\begin{eqnarray}\label{eq:transf} 
  \xi  & = & c_{\nu}\sinh^{\nu}(s/2)\equiv 1+c_{1} \sinh^{\nu}(s/2)+ c_{2}\sinh^{(\nu+2)}(s/2) \,\cdots,\\
\nonumber
\eta & = & c_{\nu}\sinh^{\nu}(t/2)\equiv 1-c_{1} \sin^{\nu}(t/2)+ c_{2}\sin^{(\nu+2)}(t/2) -
\,\cdots;\\\nonumber  
  & & 0 \le s < \infty    \quad 0 \le t \le \pi\\\nonumber
  & & \mbox{for } \nu=2,4,6,8,10, \mbox{  with  } c_{i}=0 \quad 
       \mbox{for }i > \frac{\nu}{2}, \nonumber
\end{eqnarray}
which can be written in a  differential closed form: 
\begin{eqnarray}\label{eq:transf} 
       \frac{d \xi}{ds}
        &=& D_n \sinh(s/2),\,   \frac{d \eta}{dt}= D_n \sin(t/2),
       \, D_n=\frac{(2n+1)!}{2^{2n} n!^{2}}, n=\frac{\nu}{2}-1=0,1,2, \cdots. \nonumber
\end{eqnarray}
The transformation regularizes the singularities at the nuclei by increasing the point density in the inner region. 
The higher $\nu$, the denser the points near the nuclei to ensure a better approximation of the wave function. 
The coefficients and the details are given in refs.\,\cite{Kullie:2001}, \cite{Kullie:2003}, 
As a result of this transformation one can use a square grid over $s$ and $t$.

A high value of $\nu$ (e.g. 6, 8) is needed for grids with a large number of points, 
which in turn enable a higher convergence order $q$ for the energy 
and the full utilization of a FEM approximation of the order $p$. See below and Refs. 
\cite{Kullie:2003,Kullie:20041,Kullie:20042}. 

Because of axial symmetry, the angular dependence is treated 
analytically by the ansatz:
\begin{eqnarray}\label{ANSATZ} 
\psi&=&
  \left(\begin{array}{c}
  \phi_{+}(s,t,\varphi)\\
  \phi_{-}(s,t,\varphi)
\end{array}\right)\!=\!       
\left( \begin{array}{c}
  \phi^1(s,t) \cdot e^{i(j_{z}-1/2) \cdot \varphi}\\
  \phi^2(s,t) \cdot e^{i(j_{z}+1/2) \cdot \varphi}\\
  i\phi^3(s,t) \cdot e^{i(j_{z}-1/2) \cdot \varphi}\\
  i\phi^4(s,t) \cdot e^{i(j_{z}+1/2) \cdot \varphi}\\
\end{array}\right)\ .
\end{eqnarray}
The wave function $\psi$ is an eigenstate of the total angular momentum, and the good 
quantum number $j_z$ is the $z$-component of the total angular momentum.

In the present FEM treatment, the definition domain of $s,t$ is subdivided into triangular elements $e$.
Each component ${k}$ of the relativistic wave function is approximated as \begin{eqnarray}\label{glob1}
\phi^{k}(s,t)= G^{{k}}(s,t) \sum_{{e}} \sum_{i}^n d^{k,e}_i N_{i}^{{k, e}}(s,t)\ ,
\end{eqnarray}
where $G^{k}(s,t)$ are global functions, and the sums run over all elements $e$ of the grid and over 
all $n$ nodal points $(s_i,t_i)$ of each element.  
$G^k(s_i,t_i)\, d^{k,e}_i$ is the value of the wave function at nodal point $i$.
The shape functions $N_i^{e}(s,t)$ are zero outside the element $e$. Inside they are complete polynomials  
of the order of $p$ in $s,t$ [22] 
, implementing a Lagrange-form interpolation.
The values of the coefficients of the polynomials are determined by the conditions  
that $N_i^{{k,e}}(s_j,t_j)=\delta_{ij}$ for all nodal points $i,j$ inside the element.
In our FEM implementation we use triangular Lagrangian-type elements with equidistant point distribution. 
The functions $G^{k}(s,t)$ account for the global behavior of the wave function, where $G_1^{k}(s,t)$ 
represents the angular momentum dependence and  $G_2^{k}(s,t)$ expresses the singular behavior at the two nuclei. 
They are given by 
\begin{eqnarray}\label{glob02}
G^{{k}}(s,t)= G^{{k}}_{1}(s,t) \cdot G_{2}(s,t)\ ,\nonumber\\ 
G_1^{{k}}(s,t)= ((\xi^{2}-1)(1-\eta^{2}))^{\frac{m_k}{2}}=R_{\perp}^{\frac{m_k}{2}}\ , \\\nonumber
\quad m_{1,3}=j_z-1/2, \quad m_{2,4}=j_z+1/2\ ,\\ 
\label{glob2}
G_2(s,t)= r_{1}^{-1+\gamma_{1,\kappa}}\cdot  r_{2}^{-1+\gamma_{2,\kappa}}\ .\\\nonumber
\end{eqnarray}
Here, $R_{\perp}$ is the (perpendicular) distance to the internuclear axis. 
For larger $Z$,  $\gamma_{l,k}$ 
becomes smaller and the singular behavior of the wavefunction is stronger, hence the convergence is less  efficient.
Indeed, as can be seen for the example in the hydrogenic atoms (Table\,\ref{tab:hydrogenicions}), 
the numerical precision decreases for higher atomic numbers.
Still, introducing the singular coordinate transformation given by Eq. (\ref{eq:transf}) guarantees 
a high convergence order, because it allows one to describe the singularity of the wave function near 
the nuclei more accurately \cite{Yang:1993,Yang:19911}. 
\section{Computational aspects}\label{RaD}

\subsection{Generalities}

We compute the lowest-energy state of energy  $\varepsilon_{1(1/2)g}$, i.e. with $j_z=1/2$, and gerade symmetry $g$. 
The notation of the corresponding nonrelativistic state is ${1\sigma_g}$. 
We abbreviate the notation of the (exact) energy in atomic units by the short-hand $E_{\rm rel}$. 

In all calculations we use the FEM polynomial order $p=10$. 
We run the calculation for different values of $\nu$ and size of the grid in order to achieve the best convergence. 
The size is defined by the size of the largest ellipse  $\xi=\xi_{\rm max}=const.$ containing the grid elements. 
The size of the grid can alternatively be defined by  $D_{\rm max}(\xi_{\rm max})$, defined as the distance 
between one of the centers to a point on the outermost ellipse $\xi_{max}$, where this distance is perpendicular 
to the line between the two centers \cite{Kullie:20043}. $D_{\rm max}$ values (given in atomic units) of 
approximately $30$ - $50$ are used. 
This should be compared to the most relevant value of the internuclear distance, $R=2$, the approximate 
equilibrium bond length of the ${\rm H}_2^+$ molecule. From this comparison we see that the space around 
the nuclei considered in the calculation is large compared to the internuclear distance. 

The largest number of grid points we were able to reasonably work with was $N=32\,761$. 
As we show below, for the system ${\rm H}_{2}^{+}$ a fractional uncertainty of the energy of the  
order of $10^{-20}$ is thereby achieved. For the relativistic shift, the absolute uncertainty is of the 
order of $10^{-22}$ atomic units, where we profit from an error cancellation concerning the non-relativistic 
energy $E_{\rm nrel}$. Thus, this is  a high-performance calculation. 
The computing time for the grid sequence up to $648/32761$ (see Table \ref{tab:nu2}) was 15 - 20 core-hours on a supercomputer. 
The code is not parallelized and runs on one core.
\begin{table}[b]
  \begin{tabular}{llll} \hline 
 $N_{e}/N$& \hspace{1cm}Relativistic, $E_{\rm rel}$  & 
  \hspace{1cm} Nonrelativistic, $E_{\rm nrel}$ & \hspace{0.3cm} rel. shift $\Delta E_{\rm rel}$ $(10^{-6})$
  \\ \hline
 8/441     & { \bf-1.102}281044470406312209 &{ \bf-1.102}2736006131289397101 &  { \bf-7.}44385727737249928\\
 32/1681   & { \bf-1.1026415}75567778026666 &{ \bf-1.1026342}090292764774058 &  { \bf-7.36653}850154926045\\
 72/3721   & { \bf-1.1026415810}12739732920 &{ \bf-1.1026342144}751026384076 &  { \bf-7.36653763}709451296\\
 128/6561  & { \bf-1.102641581032}40750588  &{ \bf-1.102634214494}7767797576 &  { \bf-7.3665376307}2611806\\
 200/10201 & { \bf-1.10264158103257}6082616 &{ \bf-1.10263421449494}53798398 &  { \bf-7.366537630702}77600\\
 288/14641 & { \bf-1.1026415810325771}38209 &{ \bf-1.1026342144949464}356152 &  { \bf-7.366537630702}59364\\
 392/19881 & { \bf-1.10264158103257716}2741 &{ \bf-1.10263421449494646}01360 &  { \bf-7.36653763070260}506\\
 512/25921 & { \bf-1.10264158103257716}3917 &{ \bf-1.102634214494946461}3087 &  { \bf-7.36653763070260}827\\
 648/32761 & { \bf-1.102641581032577164}097 &{ \bf-1.102634214494946461}4889 &  { \bf-7.36653763070260}893\\
extrpl$^{1}$&{ \bf-1.1026415810325771641}18 &{ \bf-1.10263421449494646150}95$^{}$ &  { \bf-7.3665376307026090}1\\
extrpl$^{2}$& &{ \bf-1.1026342144949464615089689454} &  { \bf-7.3665376307026090}3
   \\ \hline 
\end{tabular}  
\caption{\footnotesize  
Energies of {${\rm H}_2^+$} at $R=2$ and for $\alpha^{-1}=137.035999084$. 
All values in atomic units.
The calculations utilize $\nu=6$ and $D_{\rm max}=50$. $N_{e}, N$ are the numbers 
of the elements and grid points respectively.  
Superscript $^{1}$ indicates values extrapolated over the sequence $N_{e}$. Bold digits are sgnificant.
Superscrip $^2$ is with $\nu=2$, and $D_{\rm max}=50$;  
the relativistic shift is computed using the last relativistic energy value of column 2.
}\label{tab:nu2} 
\end{table}
\subsection{Convergence for the fully relativistic and the non-relativistic case}\label{sec:conv}
We present in Table \ref{tab:nu2} the relativistic and nonrelativistic energy 
values $E_{\rm rel}(N)$, $E_{\rm nrel}(N)$ of  {${\rm H}_2^+$} in grids of different point number 
$N$, including their extrapolations to an infinite grid \cite{Kullie:20043}.
The relativistic shift $\Delta E_{\rm rel}$ (fourth column) is the difference  $E_{\rm rel}-E_{\rm nrel}$. 

Performing an extrapolation is legitimate if the energy values exhibit a regular dependence on the grid size. 
This is the case for this problem if relevant parameters are chosen judiciously.
For the computation reported in Table \ref{tab:nu2} indeed one notices that with increasing number of grid points 
(finer subdivision) the accuracy increases and convergence to the true value is from above, not only 
for the nonrelativistic case, but also for the relativistic case. This behavior is known from previous 
treatments of the 2-spinor approach and is a major advantage of this method. 
As a guide to the eye, bold digits show the significant digits, but the actual uncertainty may be 
smaller than one unit of the last bold digit.

\begin{figure}[t]
\includegraphics[width=8.0cm]{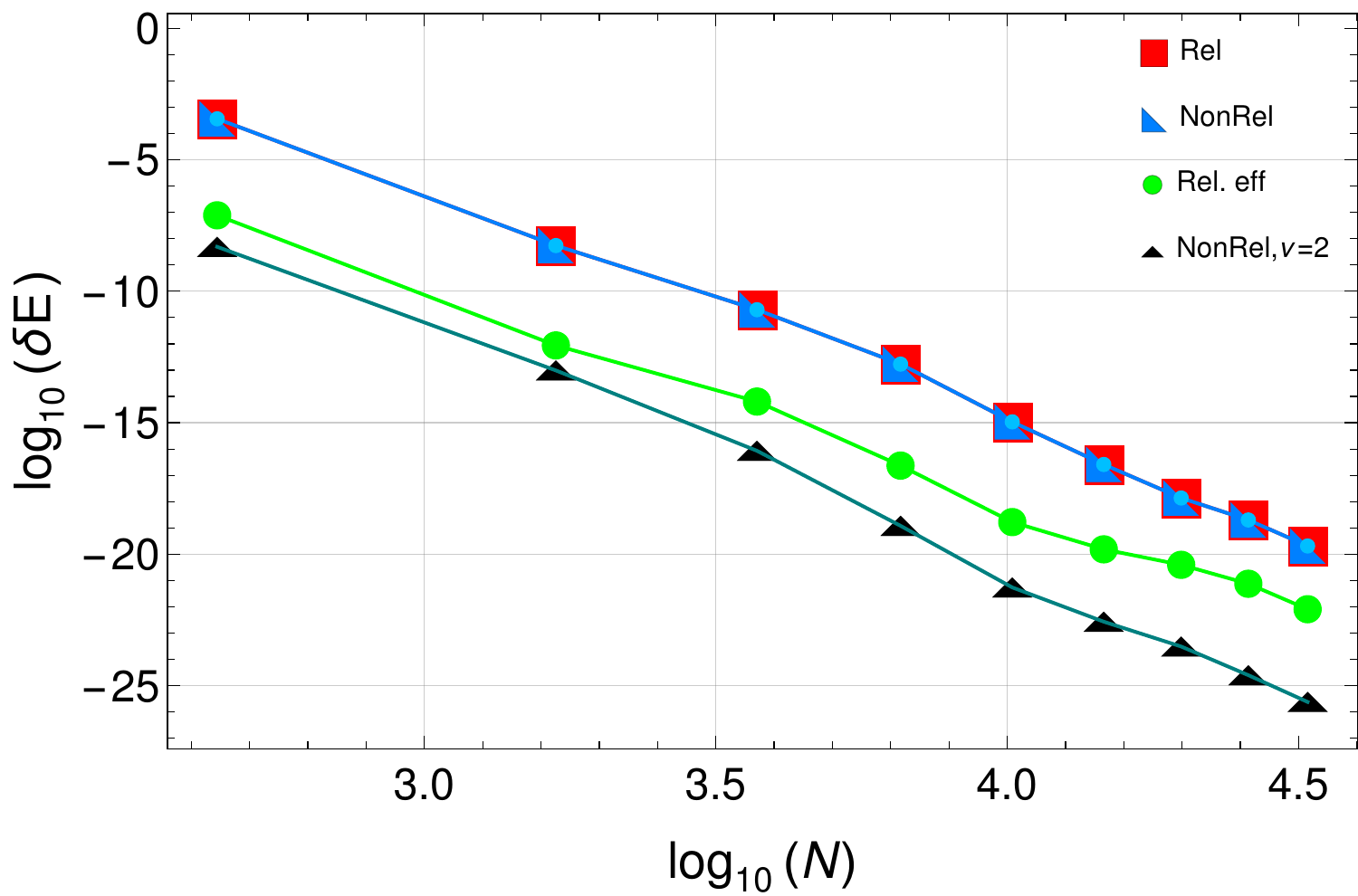}
\hspace{1cm}
\includegraphics[width=8.0cm]{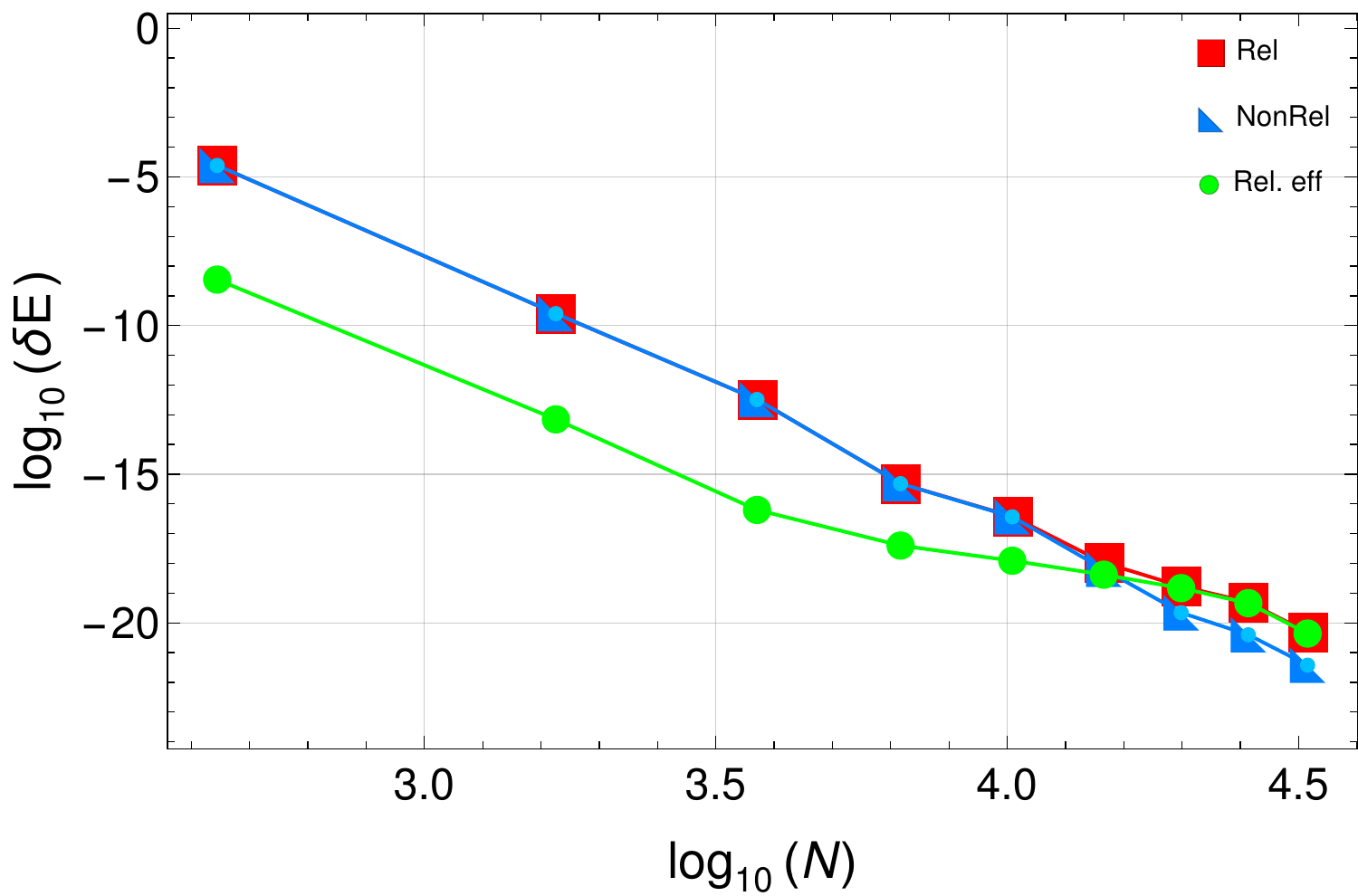}
 \caption{\label{fig:nu12} 
  \footnotesize  (Color online) 
 Convergence behavior of the relativistic (red square) and nonrelativistic (blue, left triangle)  
 energies, and of the relativistic shift (green circle), as a function of the number of grid points, $N$. 
 $\delta E=E_{}(N)-E_{}$ or $\delta E=\Delta E_{\rm rel}(N)-\Delta E_{\rm rel}$ is the deviation of 
 the energy from the extrapolated value, as given in Table \ref{tab:nu2}. 
 Two cases are considered. Left: $D_{\rm max}=50$, $\nu=6$, good convergence. 
 Right: $D_{\rm max}=70$, $\nu=4$, unsatisfactory convergence for the relativistic solution. 
 Left, black triangles:  nonrelativistic calculation with $\nu=2$ and $D_{\rm max}=50$.
}
 \end{figure}

To test the convergence and confirm the accuracy of the result, we present in Fig.\,\ref{fig:nu12} 
a log-log plot of the errors $\delta E(N)$ of the energies and of the relativistic shift computed 
for a particular grid $N$, with respect to the extrapolated value.  
As can be seen in the red and blue line of the left panel, the convergence rate in the 2-spinor 
formulation is close to that of the nonrelativistic Schr\"odinger equation \cite{Kullie:20043}. 
This is the main result of the minmax concept.

When we choose a suitably large value of $\nu$,  the values converge extremely rapidly. 
The mean convergence is approximately  $\sim N^{-8}$ considering all results from Table \ref{tab:nu2}. 
The convergence rate even appears to increase for the largest values of $N$ used. 

The accuracy in the FEM using polynomials of the order of $p$ typically scales as $\sim N^{-p}$ with grid size. 
In our case, $N$ can be made sufficiently large, but the singular behavior near the centers significantly 
reduces the efficiency of the FEM approximation. 
This can be controlled by the singular coordinates transformation given by Eq. \ref{eq:transf}, as already mentioned, 
but it is necessary to adapt the value of $\nu$ to the grid size. 
This issue is not so important in the nonrelativistic case because the wave function is finite at the nuclei.
To illustrate this we performed calculations for the same grids given in table \ref{tab:nu2}, 
but with a lower value $\nu=4$ and present the corresponding log-log plot in Fig.\,\ref{fig:nu12} (right). 

For small grids, the error evolution is similar for both the relativistic and nonrelativistic values; 
here the distribution of the grid points is balanced between inner and outer regions. 
Then, for a larger number of grid points, the error caused by the relativistic singularity 
(Eq. \ref{glob2}) becomes larger. This is because for small $\nu$ the points' density in the inner 
region is not sufficient to reproduce the singular behavior at the nuclei. 
The error of the FEM solution then does not decrease any more strongly with increasing grid size. 
In contrast, in the nonrelativistic case the error continues to decrease strongly with increasing $N$.

Returning to a larger value $\nu=6$ (Fig. \ref{fig:nu12}, left), when the wave 
function is better approximated near the charge centers, the error due to the approximate treatment 
of the relativistic singularity (Eq. \ref{glob2}) becomes smaller also for higher grid point numbers. 
Here, a suitable distribution of the grid points between the inner and outer regions of the treated domain 
is achieved \cite{Kullie:2001,Kullie:20042}.

In the relativistic shift, one profits from an error cancellation: 
the error reduces by two or more orders, as the comparison of the green and red lines 
in Fig.\,\ref{fig:nu12}, left) evidences. 
The scaling of the error of the relativistic shift with grid point number exhibits an exponent  
of approximately $q\simeq 8.5<p$. 
Determining the reason for this is beyond the scope of this work.

We take as the uncertainty of the extrapolated value the difference between the extrapolated value 
and the value computed for the most dense grid. 
Then, the achieved uncertainties are as follows.
For $E_{\rm rel}$:  $\simeq2\times10^{-20}$ atomic units, or $\simeq2\times10^{-20}$ in fractional terms. 
The same holds for the nonrelativistic energy (with the same $\nu$). 
For the relativistic shift $\Delta E_{\rm rel}=E_{\rm rel}-E_{\rm nrel}$:  $\simeq1\times10^{-22}$ 
atomic units or $\simeq1\times10^{-17}$ in fractional terms.

To show that the precise value of $D_{\rm max}$ is not crucial we present in 
Table \ref{tab:relef1} the values of the relativistic shifts  for several grid extensions $D_{\rm max}$, 
and where the grid point number is the largest one reported in Table \ref{tab:nu2}. 
As expected, the shifts stay stable over a large range of $D_{\rm max}$. 
A larger number of grid points requires larger $D_{\rm max}$ to avoid truncation errors. However, 
a larger value $\nu=6,8 ...$ condenses the points in the inner region and dilutes them in the outer region. 
To counter this effect one chooses a smaller $D_{\rm max}$ for larger $\nu$ for the same system and grids sequence. 
Note that the variation of $D_{\rm max}$ mainly affects the outer region; therefore, due to the error cancellation,  
the relativistic shift is less sensitive to $D_{\rm max}$.   

Finally, we consider the  nonrelativstic calculation. Because the wave function is not singular at 
the nuclei, $\nu=2$ is sufficient. This maintains the uniform distribution of the grid points. 
The wave function error is determined by the polynomial approximation. 
As Fig. \ref{fig:nu12} (black triangles) shows, the convergence order is approximately $q=9.5$ 
for large grid point number,  almost equal to the polynomial order $p=10$. 
Test calculations similar to those in Table \ref{tab:relef1} showed that the most accurate result 
is found for $D_{\rm max}\simeq50$, yielding the result shown in Table \ref{tab:nu2} (last line).
\begin{table}[h]
  \begin{tabular}{ccc} \hline 
 \(\begin{array}{ccc}
 D_{\rm max} & {\rm value\ for\ densest\ grid} & {\rm extrapolated\ value}\\\hline
 30 & -7.3665376307026089560\, & {\bf -7.36653763070260}899 \\
 40 & -7.3665376307026089666\, & {\bf -7.3665376307026090}4\\
 50 & -7.3665376307026089298\, & {\bf -7.3665376307026090}1 \\
 60 & -7.3665376307026088816\, & {\bf -7.36653763070260}898\\
 70 & -7.3665376307026088322\, & {\bf -7.36653763070260}894\\\hline 
\end{array}\)
\end{tabular}
\caption{\footnotesize 
Dependence of the relativistic shift $\Delta E_{\rm rel}$ (in $10^{-6}$ atomic units) 
at $R=2$ on the domain size $D_{\rm max}$. 
The second column is the result for the densest grid, with $648/32761$ elements/points. 
The third column is the extrapolated value for infinite point number. 
All evaluations were performed with $\nu=6$. Bold digits are significant}
\label{tab:relef1}
\end{table}
\subsection{\bf Test of the numerical procedure on the hydrogen-like ions}\label{sec:testofnp}
We can test our FEM procedures on the hydrogen-like ions, since the exact solution of the Dirac equation is available. 
The energy levels  
are given by (in explicit units, excluding the rest-mass energy)
\begin{equation}\label{eq:hval}
 E_{jn} = - m_{\rm e} c^{2}\left[1-
 \left(
 1+\left(\frac{Z \alpha}{(n-j-1/2)+\sqrt{(j+1/2)^{2}-(Z \alpha)^{2}}} 
 \right)^{2}
 \right)^{-1/2}
 \right]\ .
 \end{equation}
The relativistic energy shift in atomic units is 
$\Delta E_{\rm rel}=E_{jn}/(1\,{\rm a.u.})-E_{\rm nrel}=(m_{\rm e} c^2\alpha^2)^{-1} E_{jn}+Z^2/(2\,n^2)$. 
Here, we consider the ground state, $n=1, j=1/2$.
 
The FEM values were computed by setting the second center to be a dummy center with $Z_2=0$. 
In Table \ref{tab:hydrogenicions}, we show the results for various values of $Z_1$, each obtained 
with 10$^{\rm th}$-order FEM, 
with the maximum number of grid points computed being 32761, and extrapolated. 

The agreement with the exact results is excellent. 
For the case of nuclear charge $Z_1=1$ the difference of the relativistic shifts is of the order of 
$ 10^{-24}\,$a.u., or $5\times10^{-19}$ fractionally. 
This value confirms our uncertainty estimate given above 
for the two-center system $\hbox{\rm H}_2^+$.
The last two entries in the table compare the result of the energies for the relativistic and 
nonrelativistic cases. We see that our FEM is almost as accurate in the relativistic case as in 
the nonrelativistic; there is only a factor 2 loss in accuracy.

Note that unlike what is obviously the case in Eq. (\ref{eq:hval}), in the FEM computations of 
both  the hydrogenic ions and the two-center problem, varying $Z_1$ or $Z_1,Z_2$ simultaneously is 
not equivalent to varying $\alpha$ (such a variation is reported in Table \ref{tab:relativistic shift vs c} below). 
This is because our computations of the $\alpha$-dependence are based on  Eq. (\ref{eq:hhD2}) with 
fixed atomic energy unit and atomic distance unit.  
\begin{table}[h]
\begin{tabular}{llllc} \hline
 $Z$ & \hspace{1cm}$\Delta E_{\rm rel}$, exact (a.u.) & \hspace{1cm} $\Delta E_{\rm rel}$, 
 FEM (a.u.) & difference (a.u.)&  note\\ \hline
  \multicolumn{4}{c}{$\alpha^{-1}=137.0359895$}  \\\hline 
1  & -6.65659748374605054203\,$\times10^{-6}$  &{\bf -6.6565974837460505}39\,$\times10^{-6}$ & $-3.0\times10^{-24}$ &$^1$\\
2 & -1.06514068278487728906\,$\times10^{-4}$   &{\bf -1.06514068278487728}81\,$\times10^{-4}$&$-1.0\times10^{-22}$&$^1$\\
\hline  
 \multicolumn{4}{c}{$\alpha^{-1}=137.035999084$}  \\
 \hline 
1 & -6.6565965526253642790$\times10^{-6}$ & {\bf -6.6565965526253642}81$\times10^{-6}$ & \,$-2.0\times10^{-24}$&$^2$ \\ 
2 &-1.0651405337817627608$\times10^{-4}$ & {\bf -1.06514053378176276}1$\times10^{-4}$ &$2.2\times10^{-22}$&$^3$ \\
10& -6.674201689468916918\,$\times10^{-2}$ &{\bf -6.67420168946891691}3\,$\times10^{-2}$ & $-4.8\times10^{-20}$&$^3$ \\
20&-1.076523210794734716& {\bf -1.07652321079473471}3 & $-3.1\times10^{-18}$&$^3$ \\
30& -5.524906318343685119  & {\bf -5.524906318343685}09& $-3.0\times10^{-17}$&$^3$ \\
\hline\hline
 & \hspace{1.cm} $E_{\rm rel}$, exact  & \hspace{1.cm} $E_{\rm rel}$, FEM   &   difference \\ 
30 & 
$    -455.52490631834368512$ &
{\bf -455.52490631834368}34&
$-1.7\times10^{-15}$&$^3$ \\
    \hline
 & \hspace{1.cm} $E_{\rm\bf nrel}$, exact  & \hspace{1.cm} $E_{\rm\bf nrel}$, FEM   &   difference \\ 
30 & \hspace{1.5cm} $-450$ & {\bf -449.99999999999999}8309 & $-6.9\times10^{-16}$&$^3$ \\
\hline
\end{tabular}
\caption{\footnotesize  
Comparison of the extrapolated FEM numerical results for the relativistic shift of hydrogen-like ions 
with the exact result. $Z$ is the nuclear charge. 
Exact values are computed using Eq. (\ref{eq:hval}). The cases $Z=1,2$ are computed with two different 
values of $c$. The first value used is for ease of comparison with other values reported in the literature.
The next-to-last entry shows the actual relativistic energies. 
The last entry reports a nonrelativistic calculation. 
FEM parameters used: 
 Superscript $^1$: $D_{\rm max}=50$, $\nu=4$;
 Superscript $^2$: $D_{\rm max}=40$, $\nu=6$;  
 Superscript $^3$:  $D_{\rm max}=30$, $\nu=6$.
\label{tab:hydrogenicions} 
}
\end{table}
\subsection{\bf Test of the numerical procedure on an exactly known excited nonrelativistic molecular state}
It is little-known that there exist exact solutions of the nonrelativistic two-center problem for 
particular combinations of charge values $Z_1,\,Z_2$ and distances $R$ \cite{Demkov:1968}. 
These solutions are of great interest, because performing the corresponding FEM calculation allows one to 
verify the extrapolation procedure and the uncertainty estimate.
Unfortunately, these exact solutions are not the electronic ground states but excited states. 
We have treated the case ($Z_1=3$, $Z_2=2$, $R=\sqrt{15}$), whose  $4s\sigma$ state has the exact 
nonrelativistic electronic energy $E_{\rm nrel}=-(1/2)$\,a.u. 
The FEM computation 
is very cumbersome, because this state is the 21$^{\rm st}$ $j_z=1/2$ state in order of increasing energy, 
and all intermediate states have to be calculated before treating the state of interest. 
As for other calculations in the present study, the energy values are iterated until they remain stable 
at the level of $\sim 10^{-27}-10^{-30}$\,a.u.

The result is shown in Table\,\ref{tab:Z3Z2}. The difference of the extrapolated value relative 
to the exact value is $5\times10^{-20}$\,a.u. This agreement obtained for a system having substantially 
larger squared charge than the {${\rm H}_2^+$}  system, comparatively large ground state energy 
$-5.01691106677841796618$\,a.u. (significant digits, extrapolated), and using a moderate value of $D_{\rm max}$, 
gives us confidence that our uncertainty estimates for the relativistic solution of 
 {${\rm H}_2^+$}  are reasonable.
\begin{table}[t]
\begin{tabular}{ccc} \hline
 $N_e/N$& \hspace{0.0cm} $E_{\rm nrel}$, FEM (a.u.) & difference (a.u.)\\ \hline
 8/441 & -0.493598112780690785516
 & $6.4\times10^{-4}$ \\ 
 32/1681 & -0.499999685189264323528
 & $3.2\times10^{-7}$ \\ 
 72/3721 & -0.499999998831658333453
 & $1.2\times10^{-9}$ \\ 
 128/6561 &  0.499999999995025573402
 & $5.\times10^{-12}$ \\ 
 200/10201 & -0.499999999999921236179
 & $7.9\times10^{-14}$ \\ 
 288/14641 & -0.499999999999998240336
 & $1.8\times10^{-15}$ \\ 
 392/19881 &-0.499999999999999914507
 & $8.5\times10^{-17}$ \\  
 512/25921 & -0.499999999999999994099
 & $5.9\times10^{-18}$ \\ 
 648/32761 &  -0.499999999999999999158
 & $8.4\times10^{-19}$\\
 {\rm extrpl} &  -0.4999999999999999999483
 & $5.1\times10^{-20}$\\
\hline 
\end{tabular} 
\caption{FEM nonrelativistic energy of a particular excited state of the two-center problem with 
$Z_1=3, Z_2=2$ and distance  $R=\sqrt{15}$, that has the exact energy $-(1/2)$\,a.u. 
The third column shows the differences FEM minus exact. 
The FEM computations were performed with $D_{\rm max}=50, \nu=4$. $N_e/N$ is the number of the grid elements/points. 
The values are truncated.}
\label{tab:Z3Z2}
\end{table}
\vfill
\eject
\section{Results}
\subsection{Series expansion of the relativistic shift}

We have computed the FEM relativistic energy shift at $R=2\,$a.u. for a set of values of $c=\alpha^{-1}$, 
ranging from 5 to 1200. The shifts are reported in Table \ref{tab:relativistic shift vs c}. 
We fitted a series $\Delta E_{\rm rel}=\sum_{s=1}^{s_{\rm max}}{d_{2s}\alpha^{2 s}}$ to the FEM data. 
The best-fitting series is with $s_{\rm max}=6$
and is reported in Table\,\ref{tab:series expansion}, together with the standard errors of the best-fit coefficients. 
The resulting fractional deviations of the fitted values from the FEM values are approximately 
$(1-10)\times10^{-16}$, substantially larger than the uncertainties of the FEM values. 
Therefore these FEM uncertainties were not taken into account in the fit.

We also show in Table\,\ref{tab:series expansion} the comparison of our best-fit series expansion 
with what we believe is the most precise perturbation series. The coefficient for the order $\alpha^2$ 
was recently recomputed by Korobov \cite{Korobov2022privcomm}. 
It differs by $7\times10^{-12}\,\hbox{\rm a.u.}$ from the value 
in the supplemental material of Korobov (2018) \cite{Korobov:2018} (file "tmph-2017-0313-File001.dat"). 
It is now in agreement with the FEM-computed coefficient.
We recognize that a main difference between the perturbation result and FEM result is the coefficient of 
the order of $\alpha^4$. 
The uncertainty of the perturbation coefficient value as calculated by Korobov is not known.
However, our result for the $\alpha^6$-order coefficient confirms the value and uncertainty estimate of 
Mark and Becker \cite{Mark:1987}.
In total, at $R=2$ a.u. the FEM result is approximately 0.3 kHz smaller than the perturbation result.
\begin{table}[h]
\includegraphics[width=12.0cm]{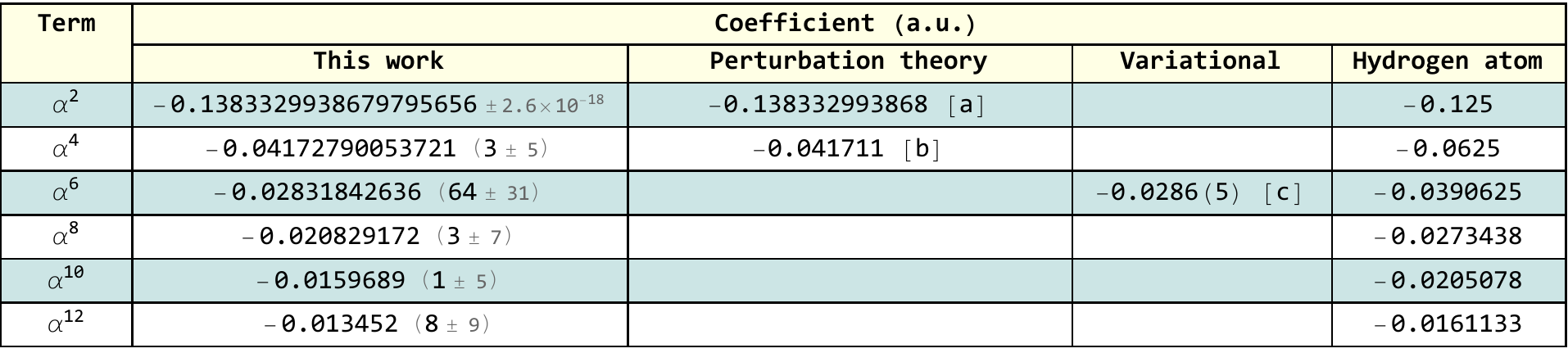}
\caption{\label{tab:series expansion}  
\footnotesize 
 Numerical series expansion of the FEM relativistic shift $\Delta E_{\rm rel}$ and comparison 
 with perturbation and variational theory results. 
 Column 2 contains the coefficients $d_{2s}$, $s=1,\ldots,6$, for $\hbox{\rm H}_2^+$ at $R=2$ a.u. 
 Columns 3 and 4 list the available best results of perturbation and variational theory, respectively. 
 Column 5 shows rounded values of the coefficients for the exact solution of the hydrogen atom. 
 The uncertainties in column 2 are for the 95\% confidence intervals. 
 The uncertainty in column 3, line 3 is one unit of the last digit. 
[a] Ref. \cite{Korobov2022privcomm}; 
[b] Ref. \cite{Korobov:2007}; 
[c] Ref. \cite{Mark:1987}.}
\end{table}
\subsection{Electronic binding energy curve}
The $R$-dependence of the Dirac energy for $\hbox{\rm H}_2^+$ has received little attention after 
the early works by Luke {\it et al.} \cite{Luke:1969} and Bishop \cite{Bishop:1977} because, after 
the experiment of Wing {\it et al.} in 1976 \cite{Wing:1976}, for several decades no precision spectroscopy 
experiments were performed that could challenge the theory. The work of Howells and Kennedy is one of 
the few exceptions \cite{Howells1990}. In the early 2000s, Korobov computed the $R$-dependent 
precise perturbation theory in connection with the new generation of experiments that had started. 
The result consists of the $\alpha^2$-coefficient \cite{Tsogbayar:2006}, whose updated values have 
been provided by Korobov for this work, and the $\alpha^4$-coefficient \cite{Korobov:2007,Korobov:2018}. 
We denote the relativistic shift computed from this data by $\Delta E_{\rm rel,Korobov}$. 

We have computed by FEM the relativistic shift for $R$-values from $R=0.05$ to $R=5.0$ 
in steps of $0.05$, see Table \ref{tab:R-dependence}. 
From the smallest to the largest $R$, the absolute uncertainties vary from  $7\times10^{-20}$ 
to $1.2\times10^{-23}$, i.e. from $7\times10^{-16}$ to $3\times10^{-18}$ in fractional terms.
\begin{table}[h]
\begin{tabular}{c|c||c|c||c|c||c|c} \hline 
$R$ &\hspace{0.5cm} $\Delta E_{\rm rel}(R)$  & $R$ &\hspace{0.5cm} $\Delta E_{\rm rel}(R)$  &$R$ & \hspace{0.25cm} $\Delta E_{\rm rel}(R)$ & $R$ & \hspace{0.25cm} $\Delta E_{\rm rel}(R)$  \\ 
\hline
0.05 &  -102.205220874297353 & 1.30 &   -12.448456768274240 & 2.55 &    -5.97244880905323561 & 3.80 &   -5.28072339944993510 \\
0.10 &  -94.2070961513458068 & 1.35 &   -11.840587433312817 & 2.60 &    -5.89682181016418189 & 3.85 &   -5.28498620517162691 \\
0.15 &  -85.3131813342957982 & 1.40 &   -11.287362392410170 & 2.65 &    -5.82730252634455702 & 3.90 &   -5.29060881471213234 \\
0.20 &  -76.6282196861760657 & 1.45 &   -10.783014143550542 & 2.70 &    -5.76349219079285695 & 3.95 &   -5.29751088104089074 \\
0.25 &  -68.6084308516448949 & 1.50 &   -10.322491660645794 & 2.75 &    -5.70502398629843855 & 4.00 &   -5.30561552816506309 \\
0.30 &  -61.4076071297781014 & 1.55 &   -9.9013599726656259 & 2.80 &    -5.65155994593625421 & 4.05 &   -5.31484911635246100 \\
0.35 &  -55.0368465872734128 & 1.60 &   -9.5157152038954672 & 2.85 &    -5.60278818603095446 & 4.10 &   -5.32514103071051427 \\
0.40 &  -49.4432261483497075 & 1.65 &   -9.1621125180967957 & 2.90 &    -5.55842043205063507 & 4.15 &   -5.33642349106801577 \\
0.45 &  -44.5490384746501491 & 1.70 &   -8.8375048531073833 & 2.95 &    -5.51818980322643993 & 4.20 &   -5.34863138128002584 \\
0.50 &  -40.2710819788872559 & 1.75 &   -8.5391906977447463 & 3.00 &    -5.48184882610666476 & 4.25 &   -5.36170209623029187 \\
0.55 &  -36.5297231045446188 & 1.80 &   -8.2647694632039667 & 3.05 &    -5.44916765105193533 & 4.30 &   -5.37557540494242218 \\
0.60 &  -33.2527185045808470 & 1.85 &   -8.0121032479359465 & 3.10 &    -5.41993244895295341 & 4.35 &   -5.39019332833313161 \\
0.65 &  -30.3763935052245251 & 1.90 &   -7.7792839978640807 & 3.15 &    -5.39394396828117659 & 4.40 &   -5.40550003025016080 \\
0.70 &  -27.8455359036944480 & 1.95 &   -7.5646052307100206 & 3.20 &    -5.37101623503049416 & 4.45 &   -5.42144172053575057 \\
0.75 &  -25.6127144763335920 & 2.00 &   -7.3665376307026090 & 3.25 &    -5.35097538022926582 & 4.50 &   -5.43796656894539199 \\
0.80 &  -23.6373874480319195 & 2.05 &   -7.1837079333992589 & 3.30 &    -5.33365858154336776 & 4.55 &   -5.45502462883235896 \\
0.85 &  -21.8849832650076221 & 2.10 &   -7.0148806141308055 & 3.35 &    -5.31891310709145427 & 4.60 &   -5.47256776958247562 \\
0.90 &  -20.3260390746755839 & 2.15 &   -6.8589419712531499 & 3.40 &    -5.30659545098689629 & 4.65 &   -5.49054961685173189 \\
0.95 &  -18.9354315392755268 & 2.20 &   -6.7148862598532203 & 3.45 &    -5.29657055133523823 & 4.70 &   -5.50892549972265814 \\
1.00 &  -17.6917086864986668 & 2.25 &   -6.5818035851749282 & 3.50 &    -5.28871108247578041 & 4.75 &   -5.52765240395459001 \\
1.05 &  -16.5765189239048591 & 2.30 &   -6.4588693097269201 & 3.55 &    -5.28289681418178407 & 4.80 &   -5.54668893055877092 \\
1.10 &  -15.5741278636581426 & 2.35 &   -6.3453347653767748 & 3.60 &    -5.27901403134357336 & 4.85 &   -5.56599525898221531 \\
1.15 &  -14.6710118162609399 & 2.40 &   -6.2405190930066166 & 3.65 &    -5.27695500836772192 & 4.90 &   -5.58553311423485584 \\
1.20 &  -13.8555168671494631 & 2.45 &   -6.1438020585500913 & 3.70 &    -5.27661753314668177 & 4.95 &   -5.60526573734309895 \\
1.25 &  -13.1175733519842682 & 2.50 &   -6.0546177163079877 & 3.75 &    -5.27790447599794821 & 5.00 &   -5.62515785855980919 \\
 \hline
\end{tabular}
\caption{\footnotesize Relativistic shift $\Delta E_{\rm rel}$, in units $10^{-6}$\,a.u., 
as a function of internuclear distance. $\alpha^{-1}=137.035999084$. 
The calculation is performed with $D_{\rm max}=35, \, \nu=8$ in the range $R=0.05$ - $0.25$,  
$D_{\rm max}=40, \, \nu=8$ in the range $R=0.30$ - $1.95$ and with $D_{\rm max}=40,\,\nu=6$ for $R=2.0$ - $5.0$, 
and in  addition extrapolated to infinite grid point density.   
}\label{tab:R-dependence}  
\end{table}
\section{Discussion and Conclusion}

\subsection{Comparison with other work}
Historically, the first numerical solution of the Dirac equation for the two-center problem was undertaken 
by Pavlik and Blinder \cite{Pavlik1967}, followed by Luke {\it et al.} \cite{Luke:1969} 
and by M{\"u}ller {\it et al.} \cite{Mueller:1973}. 
Two decades later, the precision had improved by five orders with the work of Yang  {\it et al.} \cite{Yang:1991}. 
Another one order of improvement followed in the next decade, reported by Kullie and Kolb \cite{Kullie:2001}. 
In the two decades from that work until the present work no other independent result was published with 
comparable or lower uncertainty, to the best of our knowledge. 
Several studies of the Dirac equation were undertaken, but they were not aimed towards record precision 
for the $\hbox{\rm H}_2^+$ system. 
Table\,\ref{tab:other} presents some results on the total energy, that were published in the last 25 years. 
For earlier listings of results on the Dirac equation for $\hbox{\rm H}_2^+$, 
see e.g. Rutkowski and Rutkowska \cite{Rutkowski:1987}, Sundholm \cite{Sundholm:1987}, 
and Mironova {\it et al.} \cite{Mironova:2015}. 
Table\,\ref{tab:comparison} instead focuses on the relativistic shift, going back to the earliest works. 
We see that the present result is not or only marginally in agreement with 
Refs. \cite{Kullie:2001,Korobov:2007,Mironova:2015}, perhaps because their uncertainties were underestimated. 
\begin{table}
\begin{tabular}{lll} \hline 
Reference &\hspace{1cm}  $E_{\rm rel}$ \\ 
\hline
 This work                                        & { -1.10264158103360758005}\\ 
 Mironova et al. 2015 \cite{Mironova:2015}        & { -1.102641581033}0\\
 Tupitsyn et al. 2014 \cite{Tupitsyn:2014}        & { -1.1026415810330} \\
 Fillion-Gourdeau et al. 2012 \cite{Fillion:2012} & { -1.102641580782}\\
 Artemyev et al. 2010 \cite{Artemyev:2010}        & { -1.1026409}\\
 Ishikawa et al. 2008 \cite{Ishikawa:2008}        & { -1.102641581033}598\\
 Kullie and Kolb 2001 \cite{Kullie:2001}          & { -1.102641581033}58 \\
 Parpia and Mohanty  1995  \cite{Paripa:1995}     & { -1.10264158}01\\
 Sundholm  1994  \cite{Sundholm:1994}             & { -1.102641581}\\ 
Yang et al. 1991 \cite{Yang:1991}                 & { -1.1026415810336}\\ 
\hline 
\end{tabular}  
\caption{\footnotesize 
Comparison of present and literature values for the ground-state energy of the ${\rm H}_2^{+}$ 
molecular ion at $R= 2$. This work: FEM result computed with $\nu=6, D_{\rm max}=40\,$. 
All results were obtained for  $\alpha^{-1}= 137.0359895$.
}\label{tab:other} 
\end{table}
\begin{table}
\includegraphics[width=12.0cm]{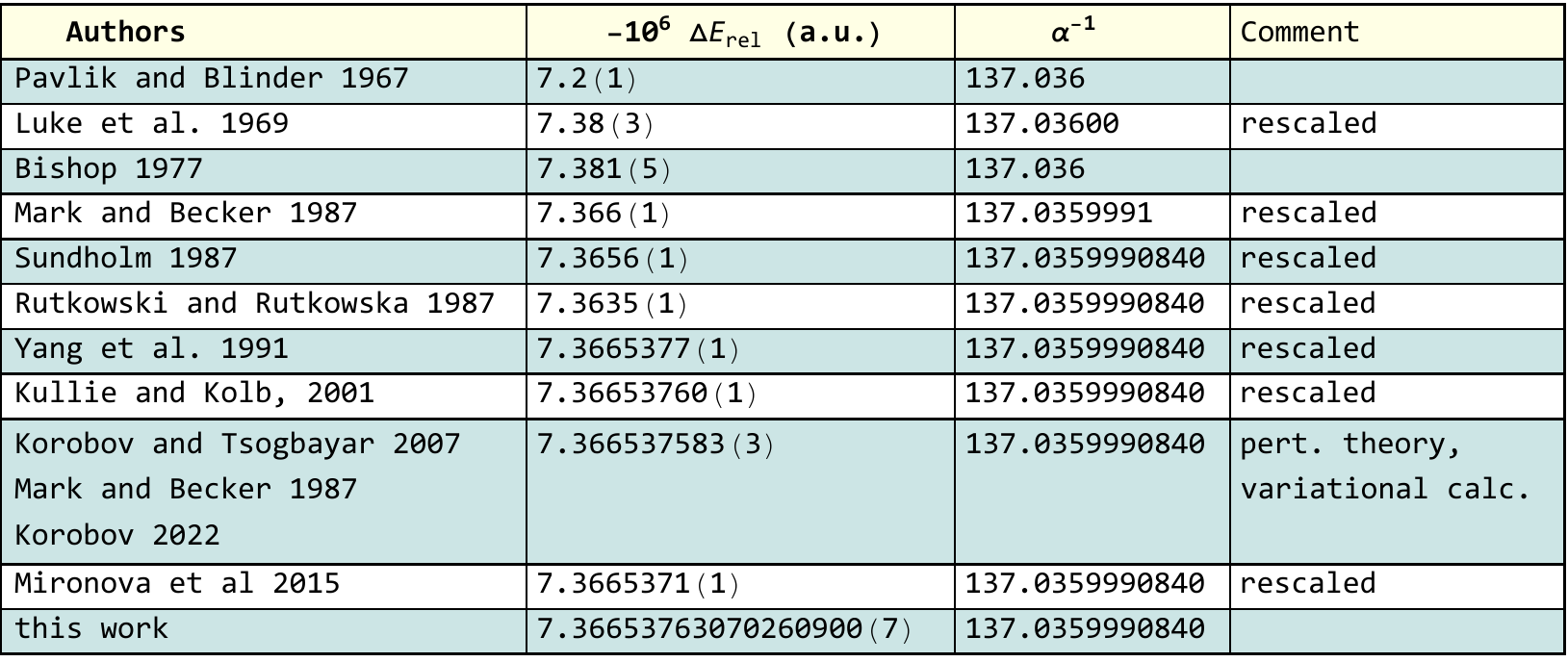}
\caption{\label{tab:comparison}  
\footnotesize 
Some results on the relativistic shift at $R=2\,$a.u. 
The value of $\alpha$ used for the computation is indicated. "Rescaled" means that 
the original computation was made for a different value of $\alpha$. 
We rescaled the original value of the shift by $(\alpha/\alpha_{\hbox{\rm \tiny orig}})^2$ 
in order to allow comparison of the shift values. This rescaling is adequate, given the moderate 
number of significant digits in the quoted works.  
The uncertainty indicated in parentheses is either taken from a statement of the authors, or, in absence, 
the unit of the last digit of the reported shift value is assumed. 
Notation: "7.38(3)" means an uncertainty of 0.03. }
\end{table}

\begin{figure}
\includegraphics[width=9.0cm]{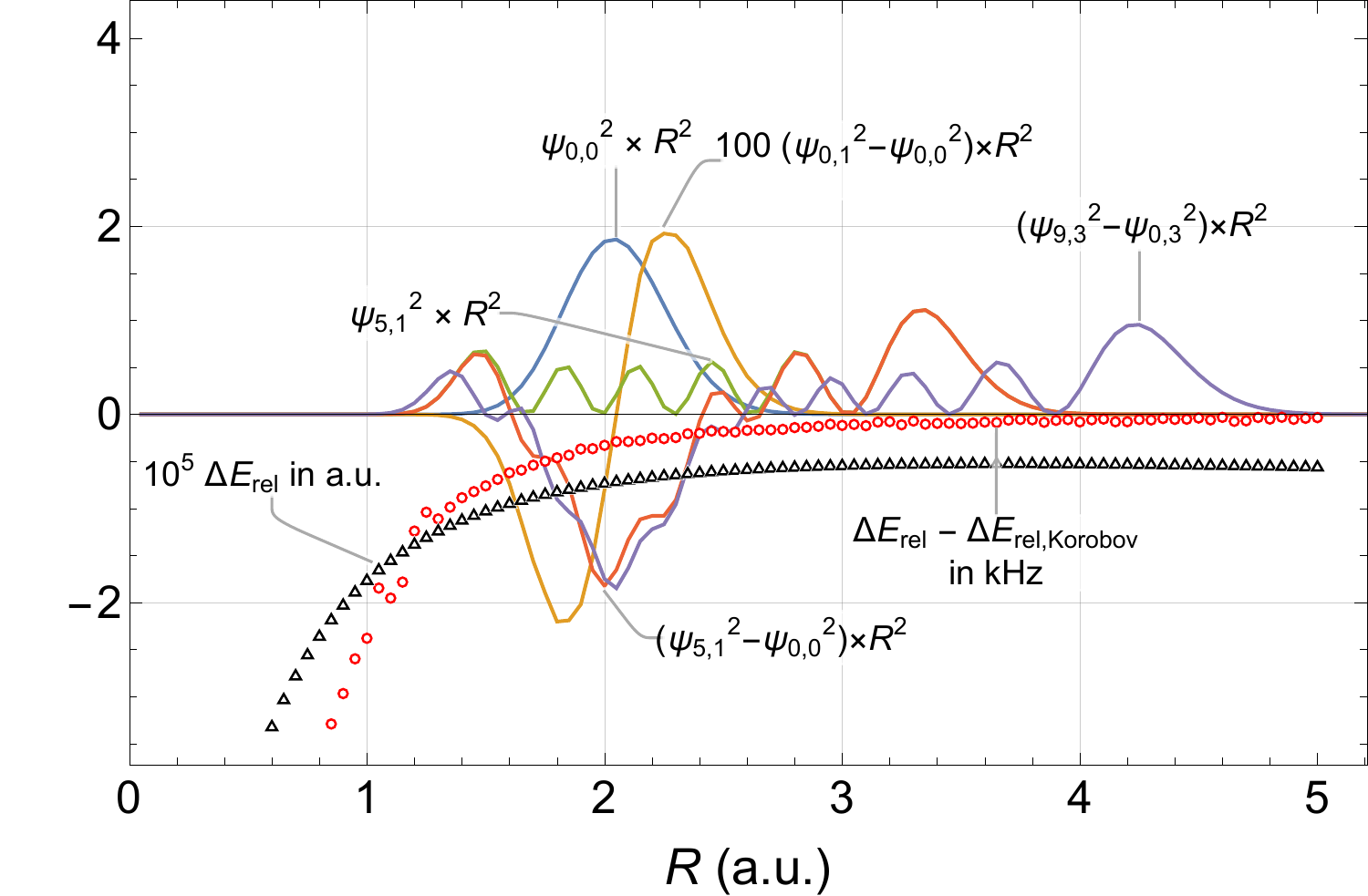}
\caption{\label{fig:wfct}
\footnotesize  (Color online) 
Black triangles: FEM relativistic energy shift $\Delta E_{\rm rel}$ for the two-center 
problem as a function of nuclear distance $R$. Red circles: difference between FEM and perturbation theory. 
The units of these two quantities are stated in the labels. 
The values of a selection of red circle points is given in column 3 of Table \ref{tab:relvsper}. 
Other lines: the nuclear radial probability densities and density differences for some relevant 
rotational-vibrational levels and transitions of the ${\rm HD}^+$ molecule.}
 \end{figure}

Figure \ref{fig:wfct} plots in red the difference between the FEM result and $E_{\rm rel,Korobov}$ 
for some of the available values of $R$. The same value of $c$ is used. 
The difference is less than 1.5 kHz in magnitude for the  range of $R$ values where the considered 
nuclear wavefunctions have significant probability. Note that the uncertainty of the FEM values is 
estimated to be of the order of $10^{-22}$\,a.u.
$\simeq10^{-9}$ kHz, completely negligible on the scale of the plot. 
We have not analyzed the origin of the visible small deviations of the values from 
a smooth dependence, especially near $R=1$, as they do not impact the following discussion. 
For concreteness, Table \ref{tab:relvsper} compares our values of the relativistic shift with 
the previous result, for selected values of internuclear distance.
\begin{table}
\begin{tabular}{c|c|c} \hline
$R$ & $\Delta E_{\rm rel,Korobov}\ (10^{-6}$\,a.u.) 
& $\Delta E_{\rm rel}-\Delta E_{\rm rel,Korobov}$ ($10^{-14}$\,a.u.)\\ \hline
0.5 &  -40.2710800643   
&  -193.577\\
1.0 &  -17.6917083247   
&  -36.1806 \\ 
2.0 &  -7.36653763008   
&  -4.9999  \\
3.0 &  -5.48184882611   
&   -1.77565 \\
4.0 &  -5.30561552817   
&   -1.21509  \\
5.0 &  -5.62515785856   
&   -0.522041  \\
%
\hline
\end{tabular}
\caption{
\footnotesize  
Relativistic shifts: column\,2 is the perturbation results $\Delta E_{\rm rel,Korobov}$;  
column\,3 is the deviations of the FEM results, $\Delta E_{\rm rel}$, from the perturbation results. 
For the FEM values, see Table \ref{tab:R-dependence}. Note: $1\times10^{-14}$ a.u. corresponds to 66\,Hz.
}\label{tab:relvsper} 
\end{table}
\subsection{Relativistic shift of transition frequencies}
A key question is whether the highly accurate results obtained in this work affect the interpretation of 
the experiments performed so far. Those interpretations were based on the theoretical treatment of 
Korobov \cite{Korobov2021} and coworkers.

We compute the relativistic shift to any rotational-vibrational level energy as the average over the nuclear 
rotational-vibrational probability density,
\begin{equation}
\Delta E_{\rm rel}(v,L)=\int_{R_{\rm min}}^{R_{\rm max}}{\Delta E_{\rm rel}(R)(\psi_{v,L}(R))^2R^2dR}    
\end{equation}
Here, $\psi_{v,L}(R))$ is the nuclear wave function, and $v,\,L$ are the vibrational and rotational 
quantum numbers of the level. 
The relativistic shift of a transition frequency between an upper level $(v',L')$ and a lower 
level $(v,L)$ is $\Delta\nu_{\rm rel}(v,L,v',L') = \Delta E_{\rm rel}(v',L')- \Delta E_{\rm rel}(v,L)$.
This relativistic shift is not complete: there are finite-nuclear-mass corrections to it. 
However, the fixed-nuclei shifts that we treat here are the dominant ones. 
As they turn out to be small (see below), we do not need to consider their corrections.

We apply this expression to the molecule HD$^+$, since precision spectroscopic results on rovibrational 
transitions are so far available only for it. The range $\{R_{\rm min},R_{\rm max}\}$ over which we computed 
the relativistic energy shift is sufficiently large for the levels investigated experimentally so far.
The HD$^+$ wave functions $\psi_{v,L}(R)$ 
were obtained by averaging over the electronic degrees of freedom the full nonrelativistic 
three-body wave function computed by Korobov.
 
We show in Fig. \ref{fig:wfct} the differences between the squared wave functions of a few levels 
that have recently been studied experimentally. Because these differences oscillate as a function of $R$, 
and because the theory difference $\Delta E_{\rm rel}-\Delta E_{\rm rel,Korobov}$ is a slowly varying 
function over the range of $R$ values where the levels in question have a substantial probability density, 
the differences (corrections) $\delta\nu=\Delta\nu_{\rm rel}-\nu_{\rm rel,Korobov}$ turn out to be small. 
The values are reported in Table \ref{tab:corr-trans-freq}.

We conclude that the corrections $\delta\nu$ are negligible compared to today's uncertainty of 
the QED contributions, which amount to $\simeq1\times10^{-11}$ relative to the transition frequencies.
Nevertheless, we expect that in the not-too-distant future the precise results obtained here will become 
relevant, given that the QED calculations may improve and that experiments definitely have the potential 
to improve their precision by several orders.
Another application of the present results is to use the obtained wavefunctions to compute quantities 
of relevance for a more precise treatment of the QED corrections.
\begin{table}[t]
  \begin{tabular}{ccc} \hline 
 \(\begin{array}{ccc}
 {\rm transition} & {\rm correction}\ \delta\nu \,{\rm (Hz)}& {\rm fractional\ correction} \\
 \hline
 (v=0,L=0)\rightarrow(v'=0,L'=1) &0.9&7\times10^{-13}  \\
 (v=0,L=0)\rightarrow(v'=5,L'=1) &60&2\times10^{-13}  \\
 (v=0,L=3)\rightarrow(v'=9,L'=3) &110&3\times10^{-13}  \\
\hline 
\end{array}\)
\end{tabular}
\caption{Correction $\delta\nu$ to the transition frequencies of ${\rm HD}^+$ arising from 
the present treatment of the relativistic shift, {\it to be added} to the perturbation result of Korobov. 
The fractional correction is the contribution normalized to the respective transition frequency.}
\label{tab:corr-trans-freq} 
\end{table}
\begin{table}[t]
\vspace*{-0.5cm}
\begin{tabular}{c|c|c|c||c|c|c|c} 
\hline 
$c$  & $E_{\rm rel}$ (a.u.)  &   $\Delta E_{\rm rel}\,(10^{-6}$\,a.u.) 
 &$u$\,(a.u.) & $c$  & $E_{\rm rel}$ (a.u.)  &   $\Delta E_{\rm rel}\,(10^{-6}$\,a.u.)
 &$u$\,(a.u.)\\ 
 \hline
 5  &   -1.10823616628281245725 &-5601.95178786599669  &1(-20) & 160 & -1.10263961819119284065 &-5.40369624637914234  &5(-23) \\
10 &    -1.10402174575200841030 &-1387.53125706194975  &2(-21) & 170 & -1.10263900115369296245 &-4.78665874650093787  &4(-23) \\
15 &    -1.10324985455012380333 &-615.640055177342779  &9(-22) & 180 & -1.10263848407154501463 &-4.26957659855311778  &3(-23) \\
20 &    -1.10298030822228538131 &-346.093727338920758  &5(-22) & 190 & -1.10263804646585469507 &-3.83197090823356170  &3(-23) \\
25 &    -1.10285565422468955126 &-221.439729743090708  &3(-22) & 200 & -1.10263767284587354132 &-3.45835092707980879  &3(-23) \\
30 &    -1.10278796937627037902 &-153.754881323918467  &2(-22) & 250 & -1.10263642783353080771 &-2.21333858434620445  &1(-23) \\
35 &    -1.10274716721032824078 &-112.952715381780223  &1(-22) & 300 & -1.10263575153336329279 &-1.53703841683127793  &1(-23) \\
40 &    -1.10272068892299195324 &-86.4744280454926898  &1(-22) & 350 & -1.10263534374665671285 &-1.129251710251342481  &8(-24) \\
45 &    -1.10270253726390794456 &-68.3227689614840043  &8(-23) & 400 & -1.10263507907778813941 &-0.864582841677900829  &6(-24) \\
50 &    -1.10268955437077065101 &-55.3398758241904579  &7(-23) & 450 & -1.10263489762185970781 &-0.683126913246303846  &4(-24) \\
55 &    -1.10267994897141281291 &-45.7344764663523530  &5(-23) & 500 & -1.10263476782758958165 &-0.553332643120139307  &3(-24) \\
60 &    -1.10267264354692895289 &-38.4290519824923411  &4(-23) & 550 & -1.10263467179455575659 &-0.457299609295080670  &2(-24) \\
65 &    -1.10266695836994784029 &-32.7438750013797330  &3(-23) & 600 & -1.10263459875358473660 &-0.384258638275091429  &2(-24) \\
70 &    -1.10266244745636394391 &-28.2329614174833596  &3(-23) & 650 & -1.10263454191055032518 &-0.327415603863668284  &1(-24) \\
75 &    -1.10265880834615650990 &-24.5938512100493472  &2(-23) & 700 & -1.10263449680735263922 &-0.282312406177710643  &1(-24) \\
80 &    -1.10265583004409394271 &-21.6155491474821583  &2(-23) & 750 & -1.10263446042040077440 &-0.245925454312894882  &1(-24) \\
85 &    -1.10265336172953448692 &-19.1472345880263714  &2(-23) & 800 & -1.10263443064035125509 &-0.216145404793583294  &1(-24) \\
90 &    -1.10265129327838995379 &-17.0787834434932376  &1(-23) & 850 & -1.10263440595937777350 &-0.191464431311990959  &8(-25) \\
95 &    -1.10264954276284737702 &-15.3282679009164686  &1(-23) & 900 & -1.10263438527648397256 &-0.170781537511048560  &7(-25) \\
100 &   -1.10264804821164058439 &-13.8337166941238396  &1(-23) & 950 & -1.10263436777255322485 &-0.153277606763340967  &5(-25) \\
110 &   -1.10264564725880168723 &-11.4327638552257233  &1(-22) & 1000& -1.10263435282798205741 &-0.138333035595908438  &5(-25) \\
120 &   -1.10264382115409753260 &-9.60665915107108865  &9(-23) & 1050& -1.10263433996708407279 &-0.125472137611282305  &4(-25) \\
130 &   -1.10264240002530585058 &-8.18553035938906732  &7(-23) & 1100& -1.10263432881976328289 &-0.114324816821377799  &3(-25) \\
{\bf C18}& -1.10264158103257716412 &-7.36653763070260900  &7(-23) & 1150& -1.10263431909458761106 &-0.104599641149552752  &3(-25) \\
140 &   -1.10264127240938096226 &-7.05791443450075182  &6(-23) & 1200& -1.10263431055954565992 &-0.096064599198404129  &2(-25) \\
150 &   -1.10264036271043522934 &-6.14821548876783325  &5(-23) \\
 \hline
\end{tabular}
\caption{{Energy and relativistic shift at $R=2\,$a.u. for different values of $c=\alpha^{-1}$. 
Values $E_{\rm rel}$, $\Delta E_{\rm rel}$ are extrapolations to an infinitely dense grid. 
The values are significant and the last digit is rounded.
Column 4 lists the estimated uncertainty of the relativistic shift, $u(\Delta E_{\rm rel})$, 
where the notation ($X$) stands for $\times10^X$. ${\rm C18}=137.035999084$ is the  CODATA 2018 value. 
For $\alpha^{-1}=5$ - $100$  the calculations were performed with $D_{\rm max}=40$, $\nu=8$; 
for the remaining values $D=50, \, \nu=6$ was used. 
The uncertainty is estimated as  $u=|\Delta E_{\rm rel}-\Delta E_{\rm  rel}({\rm densest\, grid})|$.}
\label{tab:relativistic shift vs c} 
}
\end{table}

{\it Note added: } Extending Table \ref{tab:nu2}, we obtained the energies for an even larger grid,
 $N e/N = 800/40401$, using $D = 50,\,\nu = 6$, $\alpha = 137.035999084$. 
 The more precise extrapolated values are $E_{rel} = -1.10264158103257716411725$,
 $E_{nrel} =-1.10263421449494646150814$, $\Delta E_{rel} =-7.36653763070260911\times 10^{-6}$. 
 The value $\Delta E_{rel}$  has an estimated uncertainty of 
 $2.3\times 10^{-23}$ a.u.,  fractionally  $3 \times 10^{-18}$. 
%
\section{Acknowledgments.}
We are very grateful to V.I.\,Korobov for motivating the subject of this work, 
providing precise nonrelativistic results, updated values of his earlier relativistic calculations, 
and the vibrational wave functions. One of us (O.K.) thanks Prof.\,D.\,Kolb for discussions and 
Prof.\,M.\,Garcia at the Universit{\"a}t Kassel for his support. 
We also thank the computing centers at Universit{\"a}t Kassel and Universit{\"a}t D{\"u}sseldorf 
for providing resources and advice. This work was supported the European Research Council (ERC) under 
the European Union's Horizon 2020 research and innovation program (grant no. 786306, "PREMOL").
\pagebreak
\providecommand{\noopsort}[1]{}\providecommand{\singleletter}[1]{#1}%
%
\end{document}